\documentclass[aps,pra,twocolumn]{revtex4-2}

\usepackage{graphicx}
\usepackage{amsmath}
\usepackage{amssymb}
\usepackage{amsfonts}
\usepackage{bm}

\begin{document}

\title{Time-resolved Electron Momentum Spectroscopy with Ultrashort Electron Pulses: Confined Probing and Effects of Vacuum Dispersion}

\author{Pieter Hessel Harkema}
\author{Lars Bojer Madsen}
\affiliation{Department of Physics and Astronomy, Aarhus University, Ny Munkegade 120, DK-8000 Aarhus C, Denmark}

\begin{abstract}
Previous theoretical studies have shown that attosecond electron dynamics can, in principle, be captured in electron momentum spectroscopy (EMS) using ultrashort electron pulses. By including further analytical considerations on the scattering probability, we here study the effect of the finite transversal extend of the projectile electron wave packet. We find that in wave packet scattering, the target momentum distribution is probed solely in a finite spatial region. This is evident from a spatially filtering Gabor transform appearing in the scattering probability, replacing the full momentum wave function appearing in the conventional plane wave treatment. In addition, by spatially shifting the target with regard to the wave packet focus, we illustrate the influence of vacuum dispersion, i.e., the spatial broadening of the wave packet as it propagates. Our findings are significant for the possibility to correctly interpret future attosecond-EMS results and the considered effects reflect fundamental aspects of wave packet scattering. The EMS setup may, therefore, constitute a useful framework for understanding scattering with finite wave packets.
\end{abstract}

\maketitle

\section{Introduction}\label{sec:introduction}

Recent advances in electron microscopy are facilitating the study of electron dynamics on the atomic time-scale through the use of ultrashort electron pulses. Optical modulation is the most promising candidate to realize this. In that approach,  the energy profile of an isolated electron pulse is modulated through the coupling to an optical field \cite{baum_attosecond_2007, barwick_photon-induced_2009, park_photon-induced_2010, feist_quantum_2015, vanacore_attosecond_2018, kozak_all-optical_2019}. After modulation, since electrons with different energy components travel at different velocities, the electron density reshapes during propagation. This reshaping has been shown to produce pulse trains containing pulses as short as hundreds of attoseconds \cite{priebe2017attosecond, kozak_ponderomotive_2018, morimoto_diffraction_2018, morimoto_attosecond_2018, schonenberger_generation_2019, black_net_2019}. Optically modulated electron pulses have been applied experimentally in the context of attosecond diffraction \cite{morimoto_diffraction_2018, morimoto_field-induced_2024}, imaging of optical near fields \cite{nabben_attosecond_2023} and for homodyne imaging \cite{gaida_attosecond_2024}. Theoretical studies have shown that modulation affects the scattering probability even on a stationary target \cite{morimoto_coherent_2021}. Theoretical studies of the scattering of ultrashort electron pulses are still few, even for isolated pulses. Any attosecond electron pulse necessarily contains a broad momentum distribution. Thus, describing their interaction with a target must go beyond the conventional use of asymptotic plane-waves. The most straightforward way of doing this is using wave packets \cite{baum_quantum_2010, PhysRevLett.105.263201, baum_physics_2013, shao_imaging_2013, shao_energy-resolved_2017}. The correct shape of these is actively investigated, as it influences the scattering probability \cite{morimoto2025scattering}.

In this paper we consider electron momentum spectroscopy (EMS) with ultrashort electron pulses. EMS relies on the simultaneous measurement of the two scattered electrons in electron impact ionization to image the momentum density of a bound target electron \cite{mccarthy_electron_1991, coplan__1994}.  Time-resolved EMS experiments have been realized on the picosecond scale \cite{yamazaki_molecular_2015, tang_development_2018}. Theoretical work, however, shows that an EMS experiment using attosecond wave packets could, in principle, be used to probe time-dependent structures on the atomic time-scale \cite{shao_time-resolved_2013,shao_time-resolved_2018}. In Ref.~\cite{shao_time-resolved_2013} the scattering of a Gaussian electron wave packet on a time-oscillating coherent superposition of hydrogen states was considered in the first Born and plane-wave impulse approximation.

We expand on previous results~\cite{shao_time-resolved_2013} by considering additional accurate approximations to the analytical expression for the scattering probability. Our analysis demonstrates that a wave packet of finite transversal extend only probes the target momentum distribution in a finite region. In the formalism, this feature is identified by the appearance of a Gabor transform introducing spatial filtering in the scattering probability under the assumption that the spatial shape of the wave packet is constant. The momentum wave function that appears in the usual plane wave treatment of EMS is, therefore, replaced by a Gabor transform, at least in the regime where the wave packet is not sharply spatially focused. As we will discuss, the Gabor transform can be used to elucidate features in the EMS scattering probabilities. In addition,  we consider the effect of vacuum dispersion on the scattering probability. Since electrons have mass, different momentum components propagate at different velocities in vacuum which spatially broadens the wave packet as it propagates. We illustrate the effect of vacuum dispersion by placing the target outside the wave packet focus. Our results present important aspects of the scattering of finite-sized wave packets, which are essential to include in order to correctly interpret EMS scattering probabilities. The effect of vacuum dispersion is fundamental for electron wave packets with broad momentum distributions, and can therefore prove relevant for other scattering configurations with ultrashort pulses. 

This paper is organized as follows. The theory of wave packet EMS is outlined in Sec.~\ref{sec:theory}. The results are presented in Sec.~\ref{sec:results}.
Section~\ref{sec:conclusion} draws the conclusion. Atomic units are employed throughout.

\section{Theory}\label{sec:theory}

\begin{figure}[t]
    \centering
    \includegraphics[width=\linewidth]{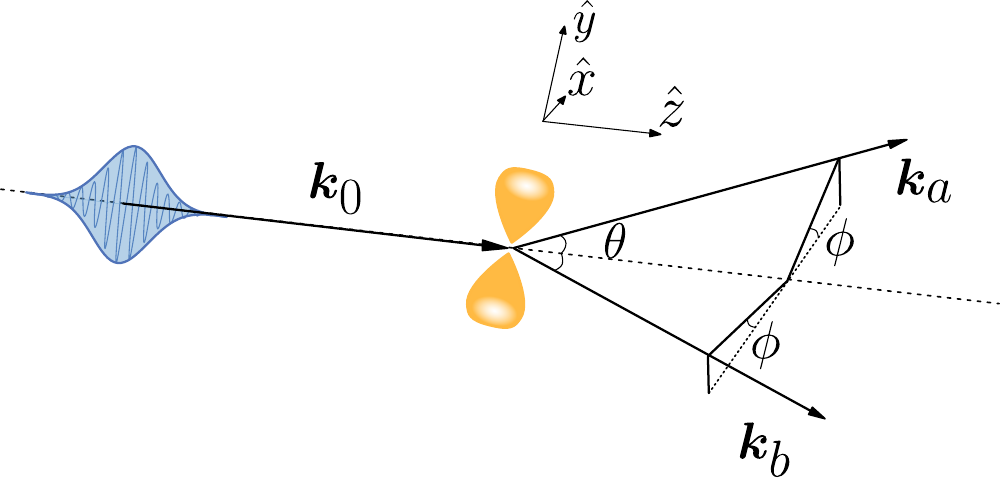}
    \caption{Wave packet EMS in the symmetric non-coplanar geometry where \(\theta = 45^\circ\) and only the angle \(\phi\) is varied. \(\bm k_0\) is the central momentum of the incident wave packet and \(\bm k_a\) and \(\bm k_b\) are the outgoing momenta.}
    \label{fig:EMSsetup}
\end{figure}

We consider the (e,2e)-electron impact ionization of a hydrogen atom as illustrated in Fig.~\ref{fig:EMSsetup}. For simplicity we consider a stationary nucleus placed at the origin. This is justified by the large mass difference between nucleus and electron. At the considered incident kinetic energy of \(10 \ \mathrm{keV}\), we apply both the first Born approximation and the plane-wave impulse approximation (PWIA) \cite{mccarthy_electron_1991, coplan__1994}. In the PWIA, the incident electron solely interacts with a single bound electron and the interaction between the continuum electrons and the nucleus is disregarded. This means that the scattering potential is given by \(V(\bm r_1, \bm r_2) = 1/|\bm r_1 - \bm r_2|\), where \(\bm r_1\) and \(\bm r_2\) are the positions of the electrons with regard to the origin. The incoming state is modeled as a coherent wave packet describing both the incident and the bound electron. In accordance with the PWIA, the incoming state is described by 
\begin{align}\label{eq:incomingstate}
    \Psi_{\mathrm{in}}(\bm r_1, \bm r_2) = \int d^3\bm{k}_i \ a_e(\bm k_i) \frac{e^{i\bm k_i \cdot (\bm r_1 - \bm b)}}{(2\pi)^{3/2}}\  \otimes \ \sum_n c_n \psi_n(\bm r_2),
\end{align}
where \(a_e\) is the momentum amplitude of the incoming electron with momentum \(\bm k_i\) and \(c_n\) the amplitude of the target wave function \(\psi_n\) with \(n\) defining a set of atomic quantum numbers. The vector \(\bm b\) denotes the impact parameter. The final state of the two continuum electrons is given by
\begin{align}\label{eq:finalstate}
    |f \rangle = |\bm k_a \rangle \otimes |\bm k_b\rangle.
\end{align}
The transition probability is found by adopting the formalism of Ref.~\cite{morimoto2024scattering} to the case of (e,2e)-scattering and is given by
\begin{align}\label{eq:TransitionProb}
    |\mathcal{A}(\bm k_a, \bm k_b)|^2 &= 4\pi^2 \sum_{n,m} c_nc_m^* \int d^3 \bm k_id^3\bm{k}_{i'} \notag \\
    \times\ & a_e(\bm k_i) a_e^*(\bm k_{i'}) \delta(E_i - E_f) \delta(E_{i'} - E_f) \notag \\
   \times\ & T_{f,i} T^{*}_{f,i'} e^{-i(\bm{k}_i - \bm{k}_{i'})\cdot \bm b}.
\end{align}
In the first Born approximation, the T-matrix elements are given as
\begin{align}\label{eq:TmatrixElements}
    T_{f,i} =& \left \langle \bm k_a, \bm k_b | V | \bm k_i, \psi_{n} \right \rangle 
    = \frac{1}{(2\pi)^3} \frac{4\pi}{\Delta^2} \tilde \psi_n(\bm q).
\end{align}
Here \(\bm \Delta = \bm k_{i} - \bm k_{a}\) is the momentum transfer to the incident electron and \(\bm q = \bm k_a + \bm k_b - \bm k_i\) is the momentum of the target electron at collision. In Eq.~(\ref{eq:TmatrixElements}), \(\tilde\psi_n(\bm q) = \frac{1}{(2\pi)^{3/2}}\int d^3 \bm r \  \psi_{n}(\bm r) e^{-i\bm q \cdot \bm r}\) is the momentum space wave function of the target. We consider EMS in the so-called symmetric non-coplanar detection geometry, as shown in Fig.~\ref{fig:EMSsetup}. In this geometry, the two outgoing electrons symmetrically share the kinetic energy, are measured at identical polar angles \(\theta = 45^\circ\), and only a single azimuthal detection angle \(\phi\) is varied \cite{mccarthy_electron_1991, coplan__1994}.
Under these conditions and with a large incident kinetic energy
\begin{align}
\label{q_formula}
    \bm q \simeq k_0 \sin{\phi}\hat{y}.
\end{align}
This means that the detection angle maps onto the momentum \(y\)-component.
In an experiment symmetric energy sharing is enforce by performing coincidence measurements.
We assume that the final energy is unresolved. This means that the observable includes integrals over the final absolute momenta.
Energy sharing is implemented by introducing an energy delta function within these integrals.
Due to the indistinguishability of the electrons, the final observable must include the effects of exchange.
By assuming that the incoming electron is unpolarized, this is done by averaging over the possible spin configurations \cite{Taylor}. The inclusion of exchange is presented in Appendix~\ref{appendix:exchange}.
The computed observable is the double differential scattering probability (DDP) given by
\begin{align}\label{eq:ExactScatProb}
    \frac{\partial^2 \mathcal{P}}{\partial \Omega_a \partial\Omega_b} (\phi) = \Big\langle \int dk_f \ k_f^3 |\mathcal{A}(k_f, \Omega_a, \Omega_b)|^2\Big\rangle_{\mathrm{spin}}.
\end{align}
The effect of exchange is largest for symmetric energy sharing \cite{coplan__1994}. However, we find it to be negligible at the considered energy.

\begin{figure*}[t]
    \centering
    \includegraphics[width=1.0\textwidth]{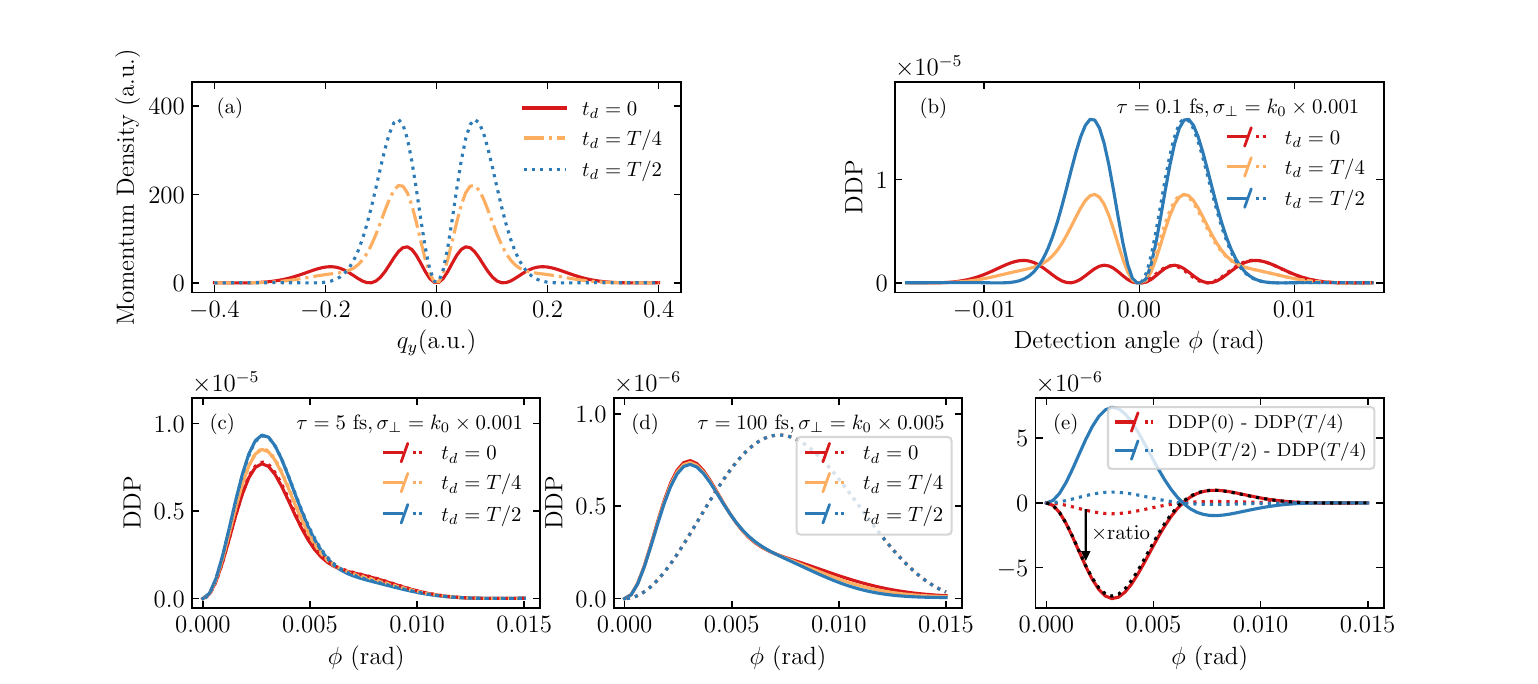}
    \caption{(a) The momentum density along the momentum space $y$-axis [see Eq.~\eqref{q_formula}] for the coherent superposition of a \(3p_y\) and \(4p_y\) state in atomic hydrogen for different values of \(t_d\). (b)-(d) Comparison between the ``exact" double differential scattering probability (DDP) from Eq.~(\ref{eq:ExactScatProb}) and Eq.~(\ref{eq:TransitionProb}) (full lines) and the approximate DDP of Eq.~(\ref{eq:ApproxScatProb}) (dotted lines) for a 10 keV Gaussian wave packet for different values of \(t_d\). In (c)-(e) only positive detection angles are considered due to the symmetry around \(\phi=0\). In (b) and (c) the wave packet is transversally narrow \(\sigma_\perp = k_0 \times 1 \ \mathrm{mrad}\) and the exact and approximate DDPs are nearly identical. The difference in the spectra originate from the overlap coefficient \(B_{3p_y, 4p_y} \approx 0.999\) for the 0.1 fs pulse and \(B_{3p_y, 4p_y} \approx 0.103\) for the 5 fs pulse. In (d), however, the projectile wave packet is transversely broad \(\sigma_\perp = k_0 \times 5 \ \mathrm{mrad}\) and the approximate DDP of Eq.~(\ref{eq:ApproxScatProb}) is no longer valid due to significant vacuum dispersion. There is a notable dependence on \(t_d\) even though \(B_{3p_y, 4p_y} \approx 0\). (e) The differences between scattering probabilities for different values of the delay parameter \(t_d\) for a 0.1 fs pulse (full lines) and a 5 fs pulse (dotted lines), respectively. For equal pulse lengths these differ only by a sign, and for different pulse lengths they are nearly the same up to a scaling factor (the ratio) as illustrated by the black dotted line. This indicates that the scattering probabilities behave similarly to temporal averages of the target momentum distribution.}
    \label{fig:ApproxScaProb}
\end{figure*}

Following Ref.~\cite{shao_time-resolved_2013}, we consider a superposition of \(3p_y\) and \(4p_y\) orbitals with time zero amplitudes \(c_n = 1/\sqrt{2}e^{-i\omega_n t_d}\).
Here, \(\omega_n\) is the energy of the hydrogen eigenstate \(\omega_n = 1/(2 n^2)\) with \(n\) now  denoting the principle atomic quantum number.
The state is imagined to be created using a finely tuned \(y\)-polarised laser \cite{Shoa_Detecting_2010}. The parameter \(t_d\) can be regarded as a controlled time delay between the creation of the target superposition and the ejection of the projectile electron. The superposition has an oscillation period of around \(T = 6.25\) fs and the momentum distribution for selected values of \(t_d\) along the \(y\)-axis is shown in Fig.~\ref{fig:ApproxScaProb} (a).  For the incident electron, we apply a Gaussian wave packet, where we assume that the longitudinal and transversal shapes separate
\begin{align}\label{eq:GaussianPulse}
    a_e(\bm k) &= \frac{1}{(2\pi\sigma_\parallel^2)^{1/4}}\exp\left(-\frac{(k_\parallel - k_0)^2}{4\sigma_\parallel^2} \right) \notag \\
    & \quad \times \frac{1}{(2\pi \sigma_\perp^2)^{1/2}} \exp\left(-\frac{k_\perp^2}{4\sigma_\perp^2}\right), 
\end{align}
where \(\sigma_\parallel\) is determined by the temporal length and \(\sigma_\perp\) by the angular spread of the pulse. For a Gaussian wave packet the temporal pulse length \(\tau\), as usually given as the full width at half maximum (FWHM), is related to the length \(\sigma_\parallel\) by \(\tau = 2\sqrt{2\ln(2)}/(2k_0\sigma_\parallel)\). For a discussion of wave packet shapes see Ref.~\cite{morimoto2025scattering}. For the pulse widths \(\sigma_\perp\) considered here, our pulse shape is nearly identical to the so-called \(k_\parallel\)-Gauss in Ref.~\cite{morimoto2025scattering}.

The full lines in Fig.~\ref{fig:ApproxScaProb} (b) show the EMS spectrum for a 100 as (FWHM), 10 keV electron pulse.  Similar to Ref.~\cite{shao_time-resolved_2013}, we use a width of \(\sigma_\perp = k_0\times 1 \ \mathrm{mrad}\). The \(\phi\)-dependent scattering probability closely mirrors the momentum distribution along the \(y\)-axis shown in Fig.~\ref{fig:ApproxScaProb} (a). Unlike Ref.~\cite{shao_time-resolved_2013}, we do not integrate over detector windows. Therefore, we find a more profound zero probability dip at \(\phi=0\) in comparison with Fig.~2 in Ref.~\cite{shao_time-resolved_2013}.

\subsection{Analysis of the scattering probability expression}

Equations (\ref{eq:TransitionProb}) and (\ref{eq:ExactScatProb}) are exact under the first Born and PWIA approximations. At sufficiently high energies, where the central momentum is large compared to the widths of the wave packet, a number of additional approximations may be applied. These allow us to gain further understanding of the relation between the shape of the wave packet and the differential scattering probability. The approximations we apply are similar to those considered in Ref.~\cite{shao_imaging_2013}. In the following, we disregard exchange effects.
Returning to Eq.~(\ref{eq:TransitionProb}), we consider the energy-conserving delta function
\begin{align}
    \delta(E_{i'} - E_i) = \delta\left(\frac{k_{i'}^2}{2} - \frac{k_{i}^2}{2} - (\omega_m - \omega_n)\right).
\end{align}
The initial momentum can be written as \(\bm k_i = k_0\hat{\bm z} + \bm {\kappa}_i\) and \(\bm k_{i'} = k_0\hat{\bm z} + \bm {\kappa}_{i'}\). Thus
\begin{align}
    k_{i'}^2 - k_i^2 = 2k_0\hat{\bm z} \cdot (\bm \kappa_i - \bm \kappa_{i'} \bm ) + \kappa_{i'}^2 - \kappa_{i}^2.
\end{align}
For a momentum distribution narrow around the central momentum \(k_0\hat{\bm z}\) with \(\kappa_i \ll k_0\), one may neglect the \(\kappa_{i'}^2 - \kappa_{i}^2\) term. Thus
\begin{align}
    \delta(E_{i'} - E_i) \simeq \frac{1}{k_0}\delta\left( k_{i', \parallel} - k_{i, \parallel} - \frac{\omega_m - \omega_n}{k_0}\right). 
\end{align}
This delta function now only restrict the coupling between longitudinal momentum components in the momentum integrals and, thus, leaves the transversal components free to interfere.
Effectively, this approximation disregards the effect of vacuum dispersion, that is, the broadening of a wave packet as it propagates through space. This is evident from the fact that the effect of vacuum dispersion originates from the time-energy phase, which for a free plane wave is given by \(e^{-ik^2t/2}\).
In the scattering amplitude, however, the time-energy phase is represented by an energy delta functions with time integrated out.
The remaining delta function enforces overall energy conservation. If the kinetic energy of the incident electron is large compared to both the longitudinal and transversal extend of the wave packet as well as the ionization energy of the target, this delta function may be approximated as
\begin{align}
    \delta(E_f - E_i) = \delta\left(\frac{2k_f^2}{2} - \frac{k_i^2}{2} + \omega_n\right) \simeq \frac{1}{\sqrt{2}k_0}\delta\left(k_f - \frac{k_0}{\sqrt{2}}\right).
\end{align}
The two approximated delta functions can be evaluated by integrating over \(k_{i', \parallel}\) and \(k_f\). This leaves
\begin{align}\label{eq:cohNote_ScatteringProb}
    \frac{\partial^2 \mathcal{P}}{\partial\Omega_a \partial \Omega_b} (\phi) &\simeq
    4\pi^2 k_0  \sum_{n,m} c_n c_m^* \int d^3\bm k_i \int d^2\bm k_{i', \perp} \notag \\ 
    \times \ & a_e(k_{i, \parallel},\bm k_{i, \perp})a_e^*(k_{i, \parallel} + \Delta k, \bm k_{i', \perp}) \notag \\
    \times \ &T_{fi}T^* _{fi'} \times e^{-i(\bm k_i - \bm k_{i'}) \cdot \bm b},
\end{align}
where \(\Delta k = (\omega_m - \omega_n)/k_0\). The T-matrix elements are given in Eq.~(\ref{eq:TmatrixElements}). Since the energy exchange is large under symmetric energy sharing the factor \(1/\Delta^2\) is nearly constant in energy. This is referred to as the EMS condition \cite{coplan__1994}. This latter factor can be evaluated simply in the central momentum \(k_0 \hat{\bm z}\) which gives \(\Delta^2 = k_0^2/2\). With this approximation the two transversal momentum integrals decouple. In addition, by assuming that the transversal and longitudinal shapes of the wave packet separate, as in Eq.~(\ref{eq:GaussianPulse}), we find the following approximate expression for the double differential scattering probability
\begin{align}\label{eq:ApproxScatProb}
    \frac{\partial^2 \mathcal{P}}{\partial \Omega_a \partial\Omega_b} (\phi) & \simeq 4\pi^2 \left((2\pi)^{-3}\frac{4\pi}{\Delta^2}\right)^2 k_0  \notag \\
    & \quad \times \sum_{n,m} c_n c_m^* B_{nm} G_n(q\hat{y}, \bm b_\perp)G_m^*(q\hat{y}, \bm b_\perp).
\end{align}
The transform \(G_n(q, \bm b_\perp)\) is called a Gabor transform and, in this case, it is the Fourier transform of the wave function times a cylindrical Gaussian window centered on the impact parameter accounting for spatial filtering
\begin{align}\label{eq:GaborTransApproxDerivation}
    G_n[q\hat{\bm y}-(k_{i, \parallel} - k_0)\hat{\bm z}, \bm b_\perp] = \frac{1}{(2\pi)^{3/2}}\int d^3\bm r \ \psi_n(\bm r) \frac{\sqrt{2\pi}}{\sigma_{r, \perp}} \notag \\
     \times \exp\left(-\frac{(\bm r_\perp - \bm b_\perp)^2}{4\sigma_{r, \perp}^2}\right) \ e^{-iq\bm{\hat y} \cdot \bm r} e^{i(k_{i, \parallel} - k_0)\bm{\hat z} \cdot \bm r},
\end{align}
where \(\sigma_{r, \perp}=1/(2\sigma_\perp)\).
The cylindrical Gaussians emerge from the decoupled transversal momentum integrals from Eq.~(\ref{eq:cohNote_ScatteringProb}), since these have been reduced to two two-dimensional Fourier transforms of the Gaussian momentum amplitudes \(a_\perp(\bm k_{i, \perp})\).
Note, in Eq.~(\ref{eq:ApproxScatProb}) an additional approximation has been performed by disregarding the longitudinal dependence in the Gabor transform, that is \(G_n[q\hat{\bm y} - (k_{i, \parallel} - k_0)\hat{\bm z}, \bm b_\perp] \simeq G_n(q\hat{\bm y}, \bm b_\perp)\). 

The overlap coefficient \(B_{nm}\) in Eq.~(\ref{eq:ApproxScatProb}) appears since also the longitudinal momentum integral has been isolated.
\begin{align}
    B_{nm}  = \int dk_{i, \parallel} \ a_{e, \parallel}(k_{i, \parallel}) a^*_{e, \parallel}(k_{i, \parallel} + \Delta k).
\end{align}
The \(B_{nm}\) factor depends on states \(n\) and \(m\) through the momentum shift \(\Delta k\). This overlap coefficient is directly related to the wave packet's ability to time-resolve the coherent oscillation of the target states \cite{robicheaux_scattering_2000, shao_imaging_2013, shao_time-resolved_2013}. If the pulse length is short compared to the target state oscillation period, that is, if \(\sigma_\parallel\) is large compared with the shift \(\Delta k\), then the overlap coefficient \(B_{nm}\) is non-zero. This results in non-zero interference terms in Eq.~(\ref{eq:ApproxScatProb}) which correspond to a temporal resolution. 

The Gabor transform introduces a spatial filtering as it extracts the momentum components in a limited spatial window using a Gaussian weighing. In this sense the formalism quantifies the intuitive expectation that a wave packet of finite transversal extend is scattered primarily around the impact parameter and thus the momentum distribution of the target state is probed only within a finite region in space. The local momentum distribution is given by \(\rho_G(q, \bm b_\perp) = |G(q, \bm b_\perp)|^2\). The Gabor transform appears due to the assumption of a separable Gaussian wave packet. A more general treatment of wave packet EMS would still find a confined momentum probing, however, with a less well-defined window function appearing in the Fourier transforms. Such a treatment is presented in Appendix \ref{appendix:general}.

\begin{figure}[t]
    \centering
    \includegraphics[width=\linewidth]{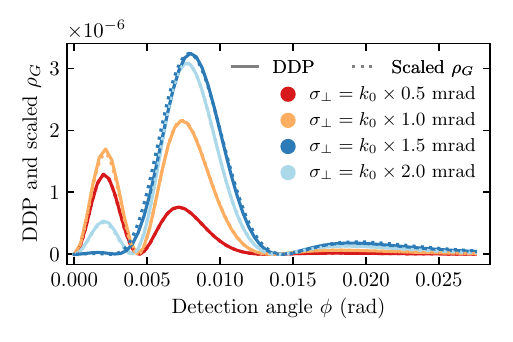}
    \caption{The double differential scattering probabilities (full lines) for a target described by \(t_d = 0\) and a pulse duration of \(\tau = 100 \ \mathrm{as}\) for different transversal widths \(\sigma_\perp\). The dotted lines, which nearly perfectly overlap with the full lines, show the momentum distributions \(\rho_G\) isolated with the cylindrical Gabor transform with an equivalent width. The momentum densities have been scaled to match the scattering probabilities. Thus in this case, the Gabor transform describes the shape of the EMS spectrum.}
    \label{fig:PulseWidthGabor}
\end{figure}

\section{Results}\label{sec:results}
\subsection{Finite probing}
We set the impact parameter \(\bm b\) to be zero. As depicted in Fig.~\ref{fig:ApproxScaProb} (b) and (c), we find that the approximate scattering probability captures the behavior of the exact probability in the regime where the wave packet is relatively narrow in momentum space (\(\sigma_\perp = k_0 \times 1 \ \mathrm{mrad}\)). For broader transversal momentum distributions, as shown in (d), the approximate expression is no longer valid. Interestingly, a dependence on \(t_d\) is apparent, even though the overlap coefficient \(B_{nm}\) is zero. We interpret this as a result of vacuum dispersion, which in this parameter regime of \(\sigma_\perp\) can not be neglected. The significant vacuum dispersion results in the wave packet being narrowly focused in space over an interval of time that is comparable to the oscillation period of the target state. Thus the momentum distribution is effectively only probed over a relatively brief period of time. 

Before we proceed, we comment on the following observation: The differences between DDPs for different values of \(t_d\) resemble the difference between averages of periodic functions. As shown in Fig.~\ref{fig:ApproxScaProb} (e), the difference \(\mathrm{DDP}(t_d=0) - \mathrm{DDP}(t_d=T/4)\) and \(\mathrm{DDP}(t_d=T/2) - \mathrm{DDP}(t_d=T/4)\) differ solely by a sign for equal pulse lengths.  In addition, for different pulse lengths the differences are nearly identical up to a scaling factor which is shown by the black dotted line. This behaviour is similar to that of an average of a periodic function.

The Gabor transformation can be used to explain key features in the wave packet EMS spectrum originating from the finite transversal extent of the wave packet. In Ref.~\cite{shao_time-resolved_2013}, it is observed that for a target state described by \(t_d = 0\) the ratio between the inner (\(\phi \sim \pm 0.002\ \mathrm{rad}\)) and outer (\(\phi = \pm 0.007 \ \mathrm{rad}\)) peaks in the EMS spectrum was inverted with regards to the corresponding peaks in the momentum distribution. This can be seen by comparing the full red lines in Fig.~\ref{fig:ApproxScaProb} (a) and (b). It was argued that this was related to the interference of transversal momentum components. With the Gabor transform a more rigorous explanation can be given. Figure~\ref{fig:PulseWidthGabor} compares the EMS spectrum for different pulse widths with momentum densities isolated with a Gabor transform of equivalent widths. The larger the width \(\sigma_\perp\), the closer to the nucleus the momentum is probed. The figure shows that the inversion of the ratio between the peaks originates from the probed momentum densities. The closer to the nucleus the momentum is probed, the more inverted the ratio becomes. This is consistent with the notion that electrons close to the nucleus have a higher momentum. The agreement between the scattering probability and the momentum density becomes less for larger values of \(\sigma_\perp\). This is consistent with what has already been discussed regarding the validity of the approximations leading to Eq.~(\ref{eq:ApproxScatProb}). 
\begin{figure}[t]
    \centering
    \includegraphics[width=\linewidth]{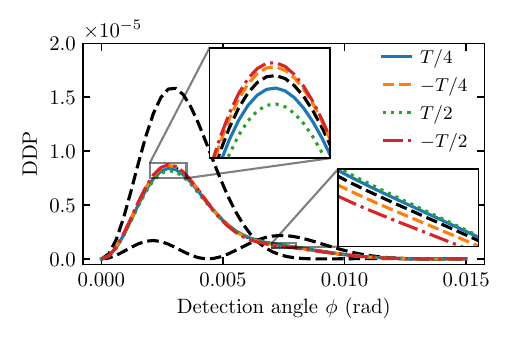}
    \caption{The EMS double differential scattering probability (DDP) of 100 as wave packet for different shifts \(t_p\) of the wave packet focus.   The target wave function is categorized by the delay parameter \(t_d = T/4\). A notable difference between positive and negative shifts is observed.  The dashed black lines represent results for a focused wave packet for \(t_d = 0,\ T/4, \ T/2\) respectively (as given in Fig.~\ref{fig:ApproxScaProb} (b)) and serve as a reference.}
    \label{fig:OutOfFocus}
\end{figure}

\subsection{Influence of vacuum dispersion}

The effect of vacuum dispersion can lead to a temporal resolution even if the wave packet has a large temporal length, as illustrated in Fig.~\ref{fig:ApproxScaProb} (d). Here, even though \(\sigma_\parallel\) is small compared to the momentum shift \(\Delta k\), the sharp transversal narrowing of the wave packet leads to a dependence on the delay parameter \(t_d\). This implies than a more meaningful definition of the temporal pulse length should be based on the full three-dimensional electron density rather than simply on the longitudinal width \(\sigma_\parallel\). Here we show that the influence of vacuum dispersion becomes apparent in the case of a non-zero longitudinal impact parameter even if the wave packet is not sharply focused (\(\sigma_\perp = k_0 \times 1 \ \mathrm{mrad}\)). A longitudinal impact parameter is equivalent to a shift of the wave packet focus. The focus is the point in time and space where the electron density is most concentrated, and is well-defined for the Gaussian wave packets applied here. The focus is shifted by including the following two phases
\begin{align}
    e^{-ik_{i, \parallel}k_0 t_p} \times e^{i k_i^2t_p}.
\end{align}
The first phase shifts the focus forward spatially by \(k_0t_p\). The second results in a temporal shift of \(t_p\). Including both shifts is thus equivalent to moving the electron source relative to the target combined with delaying projectile emission. Thereby, the moment of overlap between the projectile wave packet and the target wave function is not changed. Figure~\ref{fig:OutOfFocus} shows that the sign of the shift, that is, whether the wave packet focus is placed before or after the target, influences the EMS spectrum.  If the focus is located before the position of the target (\(t_p < 0\)) the scattering probability is slightly larger around \(\phi \simeq\pm 0.003\) and slightly smaller around \(\phi \simeq\pm 0.0075\) when compared the case \(t_p>0\). We note that the asymmetry persists for larger shifts, e.g., for \(t_p = \pm 20T/40,\ 21T/4\), where the total scattering probability is also reduced. 

We propose the following explanation: When the wave packet focus is located away from the origin, the spatial electron density of the wave packet is either slightly higher as it propagates towards the origin or away from the origin. This asymmetry translates to a difference in how intensely the target state is probed just before and just after the wave packet crosses the origin.
\begin{figure}[t]
    \centering
    \includegraphics[width=\linewidth]{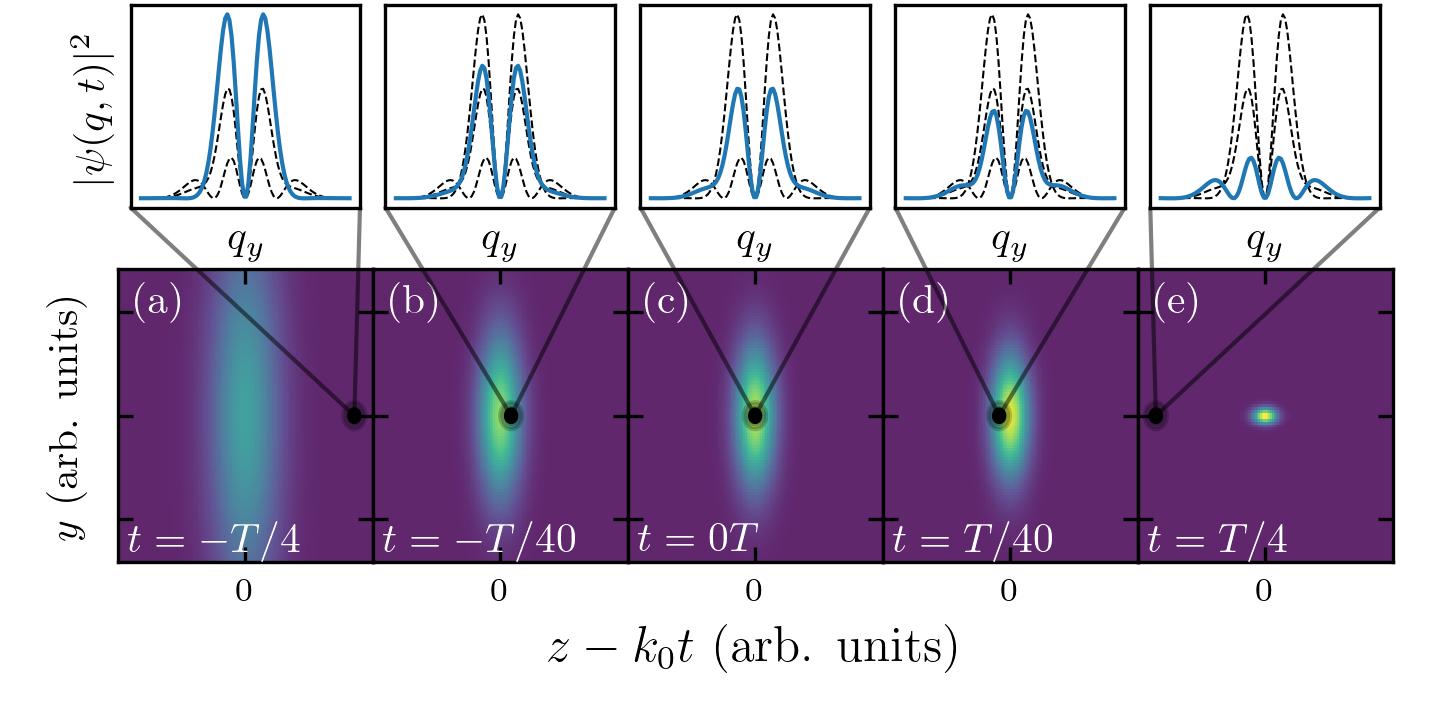}
    \caption{The propagation of a Gaussian wave packet across the origin. The wave packet focus is shifted by \(t_p = T/4\) and is shown in (e). As the wave packet crosses the target atom  (black dot), which is defined by \(t_d = T/4\), located at the origin the projectile electron density changes. When propagating away from the origin, shown in (d), the electron density is slightly higher compared to when it moved towards the origin (b). The target momentum density (shown on the insert) is therefore probed more at \(t=T/40\) than \(t=-T/40\) effecting the shape of the resulting EMS spectrum. Note, the dispersion of the wave packet has been exaggerated for illustration and does not resemble the actual dispersion for the parameters used in Fig.~\ref{fig:OutOfFocus}.}
    \label{fig:pulsepropagation}
\end{figure}
Figure~\ref{fig:pulsepropagation} illustrates the idea. The projectile electron density is slightly higher just after the wave packet passes the origin, as shown in panel (d), in comparison to just prior of passing the origin, as shown in panel (b).  We therefore imagine that the target momentum distribution is probed more at later times \(t>0\) than \(t<0\) and expect the resulting EMS spectrum to resemble the momentum density shown on the insert in panel (d).  Comparing with the full blue line in Fig.~\ref{fig:OutOfFocus} this is exactly what we find. This model is corroborated by the following observations: First, for a target described by \(t_d=0\) there is no difference between a negative and positive shift of the focus. In this case the target momentum distribution is symmetric with respect to time inversion around \(t=0\), which is not the case for \(t_d=T/4\). Second, increasing the dispersion by increasing \(\sigma_\perp\) results in a larger difference between a positive and negative shift. Decreasing \(\sigma_\perp\) results in the opposite.

\section{Conclusion}\label{sec:conclusion}

In summary, we have shed light on aspects of electron momentum spectroscopy using attosecond electron wave packets. By analyzing the differential scattering probability, we have shown that the EMS spectrum from the scattering of a finite wave packet only probes the target-electron momentum distribution in a finite spatial region. In the formalism, this is quantified through the appearance of a spatial Gabor transform. Thus, where the scattering probability is proportional to the norm-square of the full momentum wave function of the target electron in the conventional plane wave treatment, this momentum space wave function is replaced by a Gabor transform of the target-electron wave function in the case of scattering with wave packets. 

The physics captured by the spatial Gabor transform was used to further explain important features in the attosecond EMS scattering probabilities reported in Ref.~\cite{shao_time-resolved_2013}. Some result were obtained in a regime where the effect of vacuum dispersion could not be neglected, and  we found that the effect of vacuum dispersion is fundamental for wave packet EMS.  We illustrated the dispersion dependence by including a study of scattering of a wave packet with a non-zero longitudinal impact parameter. The resulting asymmetry in the scattering signal between a negative and positive impact parameter was rationalized by the difference in the intensity with which the target momentum distribution was probed as the wave packet propagated across the target. The difference in intensity being an effect of vacuum dispersion.

As the realization of attosecond electron pulses progresses, the effects of finite wave packet size and vacuum dispersion may constitute important considerations not only for ultrafast EMS but also for other applications of ultrashort electron wave packets. The close relationship between the scattering probability and the target momentum distribution makes interpreting EMS spectra relatively straightforward. EMS may therefore prove to be a useful framework for gaining insights into the scattering of ultrashort wave packets and to elucidate the effects of, e.g., vacuum dispersion or wave packet shape.

\begin{acknowledgments}
We thank Mads Br{\o}ndum Carlsen for useful discussions. This work was supported by the Independent Research Fund Denmark (Grant-ID: 10.46540/4283-00004B).
\end{acknowledgments}

\appendix
\section{The inclusion of exchange effects}\label{appendix:exchange}
We here argue that the effect of exchange can be neglected at the considered energies under symmetric energy sharing.
The spatially symmetric initial and final state, for which the total spin \(S=0\), can be found by applying the symmetrization operator to Eq.~(\ref{eq:incomingstate}) and Eq.~(\ref{eq:finalstate}) and can be written as
\begin{subequations}
    \begin{align}\label{eq:symmetrizedWF}
    |{i; S=0}\rangle = \frac{1}{\sqrt{2}}(|{\bm k_i, \psi_n}\rangle + |{\psi_n, \bm k_i}\rangle) \\
    |{f; S=0}\rangle = \frac{1}{\sqrt{2}}(|{\bm k_a, \bm k_b}\rangle + |{\bm k_b,\bm k_a}\rangle).
    \end{align}
\end{subequations}
Where we have suppressed the summation over initial momentum and quantum numbers to simplify the notation. Similarly, the antisymmetric states, with total spin \(S=1\), can be written as
\begin{subequations}
    \begin{align}\label{eq:antisymmetrizedWF}
    |{i; S=1}\rangle = \frac{1}{\sqrt{2}}(|{\bm k_i, \psi_n}\rangle - |{\psi_n, \bm k_i}\rangle) \\
    |{f; S=1}\rangle = \frac{1}{\sqrt{2}}(|{\bm k_a, \bm k_b}\rangle - |{\bm k_b, \bm k_a}\rangle).
    \end{align}
\end{subequations}
We assume an unpolarized projectile electron. Since the interaction potential does not couple to spin it is sufficient to average over the initial spin configurations \cite{Taylor}.
\begin{align}\label{eq:ExactScatProbAppendix}
    \frac{\partial^2 \mathcal{P}}{\partial \Omega_a \partial\Omega_b} (\phi) &= \Big\langle \int dk_f \ k_f^3 |\mathcal{A}(k_f, \Omega_a, \Omega_b)|^2\Big\rangle_{\mathrm{spin}} \notag \\
    &= \frac{1}{4}\int dk_f \ k_f^3 |\mathcal{A}(k_f, \Omega_a, \Omega_b, S=0)|^2 \notag \\ & \quad  + \frac{3}{4}\int dk_f \ k_f^3 |\mathcal{A}(k_f, \Omega_a, \Omega_b,S=1)|^2.
\end{align}
The two scattering amplitudes solely differ in the momentum exchange factors which are given by
\begin{align}
    \frac{1}{\Delta^2_\mathrm{sym}} = \frac{1}{|\bm k_i - \bm k_a|^2} + \frac{1}{|\bm k_i - \bm k_b|^2}
    \intertext{and}
    \frac{1}{\Delta^2_\mathrm{antisym}} = \frac{1}{|\bm k_i - \bm k_a|^2} - \frac{1}{|\bm k_i - \bm k_b|^2}.
\end{align}
However, in the symmetric non-coplanar geometry at energies that are large compared to the momentum widths of the wave packet
\begin{align}
    \frac{4\pi}{|k_n\hat{\bm k}_i - \bm k_a|^2} \approx \frac{4\pi}{|k_n\hat{\bm k}_i - \bm k_b|^2}.
\end{align}
This means that the antisymmetric spatial channel does not contribute significantly to the scattering probability. In addition, the symmetric channel can be evaluated as \({1}/{\Delta^2_\mathrm{sym}} \approx 2/\Delta^2\), with \(\Delta\) as defined in the main text. Thus, exchange effect may be neglected at high energies under symmetric energy sharing.

\section{Scattering probability for a general wave packet}\label{appendix:general}
Here we present a more general approximate expression for the scattering amplitude. We make no assumptions regarding the shape of the incoming wave packet other than that it is well-collimated in momentum space. We therefore assume that approximations leading to Eq.~(\ref{eq:cohNote_ScatteringProb}) are still valid. Assuming again the EMS condition is satisfied, a general expression for the approximate scattering probability is given by
\begin{align}\label{eq:GenApproxScatProb}
    \frac{\partial^2 \mathcal{P}}{\partial \Omega_a \partial\Omega_b} (\phi) & \simeq 4\pi^2 \left((2\pi)^{-3}\frac{4\pi}{\Delta^2}\right)^2 k_0  \notag \\
    & \times \sum_{n,m} c_n c_m^* \int dk_{i,\parallel} \ \tilde G_n[q\hat{y} - (k_{i, \parallel} - k_0)\hat{z}, \bm b_\perp] \notag \\
    & \times \ \tilde G_m^*[q\hat{y} -(k_{i, \parallel} + \Delta k - k_0)\hat{z}, \bm b_\perp],
\end{align}
which instead of the Gabor transform includes the transform given by
\begin{align}
    & \tilde G_n[q\hat{y} - (k_{i, \parallel}-k_0)\hat{z}, \bm b_\perp] = \notag \\
    & \qquad \frac{1}{(2\pi)^{3/2}}\int d^3\bm r \ \psi_n(\bm r) e^{-i[q\hat{y} - (k_{i, \parallel}-k_0)\hat{z}] \cdot \bm r} w(\bm r -\bm b_\perp, k_{i, \parallel}).
\end{align}
This is again a three-dimensional Fourier transform with a window function, $w$. Since we did not assume that the longitudinal and transversal shape of the wave packet separate the window function now depends on the longitudinal momentum.
\begin{align}
    w(\bm r - \bm b_\perp, k_{i, \parallel}) = \int d^2 \bm k_{i, \perp} \ a_e(k_{i, \parallel}, \bm k_{i, \perp})e^{i(\bm r - \bm b_\perp) \cdot \bm k_{i, \perp}}.
\end{align}

\bibliography{ref}

\end{document}